# *Measuring the interphase between solid electrolytes and lithium metal using neutrons*

*Andrew S. Westover[1]*, Katie L. Browning[1], Antonino Cannavo[2], Ralph Gilles[3]*, Jiri Vacik[4], Mathieu Doucet[5], James F. Browning[5], Neelima Paul[3], Giovanni Ceccio[4], Vasyl Lavrentiev[4]*

**[1] Chemical Sciences Division, Oak Ridge National Laboratory, Oak Ridge, 37830, USA**

**[2] Collider Accelerator Department, Brookhaven National Laboratory, Upton, NY 11973, USA**

**[3] Heinz Maier-Leibnitz Zentrum, Technical University Munich, 85748, Garching, Germany**

**[4] Nuclear Physics Institute, Czech Academy of Sciences, CZ-25068, Rez, Czech Republic**

**[5] Neutron Scattering Division, Oak Ridge National Laboratory, Oak Ridge, 37830, USA**

***Correspondence to westoveras@ornl.gov; ralph.gilles@frm2.tum.de**

## Abstract

Interfaces are the key to next-generation high-energy batteries including solid-state Li metal batteries. In solid-state batteries, the buried nature of solid-solid electrolyte-electrode interfaces makes studying them difficult. Neutrons have significant potential to non-destructively probe these buried solid-solid interfaces. This work presents a comparative study using both neutron depth profiling (NDP) and neutron reflectometry (NR) to study a model lithium metal-lithium phosphorus oxynitride (LiPON) solid electrolyte system. In the NDP data, no distinct interphase is observed at the interface. NR shows a difference between electrodeposited, and vapor deposited LiPON-Li interfaces but finds both are gradient interphases that are less than 30 nm thick. Additional simulations of the LiPON-$Li_2O$-Li system demonstrate that NDP has an excellent resolution in the 50 nm – 1 μm regime while NR has an ideal resolution from 0.1 – 200 nm with different sample requirements. Together NDP and NR can provide a complementary understanding of interfaces between Li metal and solid electrolytes across relevant length scales.

## TOC

## Introduction:

The performance of high energy density batteries including solid state Li metal batteries is defined by the quality of the electrode-electrolyte interfaces.[1-3] Solid-state batteries using Li metal paired with solid electrolytes are one of the most promising electrode-electrolyte combinations for next generation batteries.[2,4-7] The nature of this cell design contains solid-solid electrode-electrolyte interfaces that are buried.[3,8,9] Developing methodologies that can enable the study of these buried solid-solid interfaces is critical for optimizing these solid-state Li metal batteries and realizing their potential. Neutrons have long been effective due to their non-destructive evaluation of materials.[3,10-18] For Li metal applications both neutron depth profiling (NDP),[12,14,15,19-21] neutron reflectometry (NR),[11,13,22-24] and neutron imaging (NI)[25-27] have all been proposed to study the interfaces of solid-state batteries on various length scales. Each of these techniques have unique advantages and challenges that will make them ideal for evaluating interfaces in different situations and materials systems.

The discovery of the solid electrolyte lithium phosphorus oxynitride (LiPON) in the 90's was instrumental in the field of solid-state batteries.[28-31] This electrolyte enabled the invention and commercialization of the thin film battery with a thin sputtered cathode such as $LiCoO_2$, a LiPON electrolyte and a Li metal anode.[31] These batteries have been cycled for >10000 cycles[28,29] and have demonstrated charging current densities up to 10 mA/cm$^2$.[30] Because of the battery form factor and method of synthesis, the thin film geometry with LiPON and Li metal is an ideal model system to probe the Li/solid electrolyte interphases. Indeed, several studies have evaluated the interface between LiPON and Li metal and LiPON and cathodes.[3,8,32,33] Transmission electron microscopy (TEM) has been one of the most effective methods but is necessarily limited by difficult sample preparation such as cryogenic focused ion beam (Cryo-FIB), which may alter the interfaces.[32,33] TEM is also limited to small sample volumes that may or may not represent real life battery interfaces which can range in areal footprint from cm to m. Electroanalytical techniques have also been used to demonstrate LiPON's ability to effectively passivate Li metal anodes through formation of a less than 10 nm thick interphase.[8] These techniques are themselves limited by the indirect nature of the methods and work best in conjunction with TEM or neutron techniques. NR has also been used coupled with Coulombic titration to evaluate the interface of LiPON with Li metal confirming that on average the interphase between LiPON and Li metal is less than 10 nm thick.[3]

In this work, we build on prior NR studies[3] and compare NDP and NR datasets to identify advantages and limitations of each technique. The NDP data shows a detailed profile of the LiPON/Li metal sample stack. Model fits to the NDP data show no difference between models with and without an interphase layer, highlighting that NDP is unable to distinguish 10 nm thick interphase layers in these samples. When a 50 nm Ni layer is included in between the LiPON and Li, the 50 nm Ni layer is clearly distinguished in both the NDP data and the fitted model. Further simulations show that high Li content materials such as $Li_2O$ need around 100 nm of material to see the difference in the NDP data. To complement this data NR was also performed on a similar Li/LiPON cell stack. In this case the NR resolves both an electrodeposited Li/LiPON gradient interphase less than 10 nm thick and a vapor deposited Li/LiPON interphase that is ~30 nm thick. Further modeling data shows that the NR is very sensitive for interphase layers between 10-200 nm thick.  NDP on the other hand can handle thicker samples with generally less stringent sample requirements. Together this experimental and simulated data highlights the complimentary nature of both NDP and NR for studying interfaces in SSB. It should be emphasized that both methods gained an average result over many tens of $mm^2$ of sample size together with a deep penetration depth in the sample in comparison to classical microscopy methods.

**Results and Discussion:**

Figure 1 shows a schematic of both NDP and NR for evaluating the interfaces between solid electrolytes and Li metal. In NDP thermal neutrons (energy of about 0.025 eV) are focused on a sample. The neutrons will react with $Li_6$ in the sample emitting both α and triton radiation.[12,19,34,35] The α and triton particles are emitted from the point of origin and lose energy in passing through the sample. More energy is lost the deeper the reaction occurs in the sample. By detecting the energy and quantity of the emitted radiation it is possible to identify the $Li_6$ concentration as a function of depth at which the particles were produced, allowing the construction of a Li-depth profile. This depth profile is measured as a distribution of counts per energy channel. For the purposes of this paper data is presented in energy. NR measures the probability of a neutron of a given momentum to reflect off the surface of a sample.[36,37] Elastic neutron scattering of a cold neutron beam (energy from $5x10^{-5}$ eV to 0.025 eV) at small angles (typically smaller than a few degrees) to the sample surface leads to reflected neutrons with an intensity profile as a function of momentum and angle dependent on the chemical composition of the film as a function of depth. The change in composition as a function of depth gives rise to interference patterns (measured as reflectivity curves) as the scattered wave of the neutrons acquires a different phase for each path through the thickness of the film. The sum of the coherent scattering lengths for each isotope of each atom per unit volume obtains the average scattering length density (SLD) as a measure of the scattering power of a material. Analysis of the Kiessig fringes in the reflectivity curve provides information on layer thickness, density, roughness, and composition.

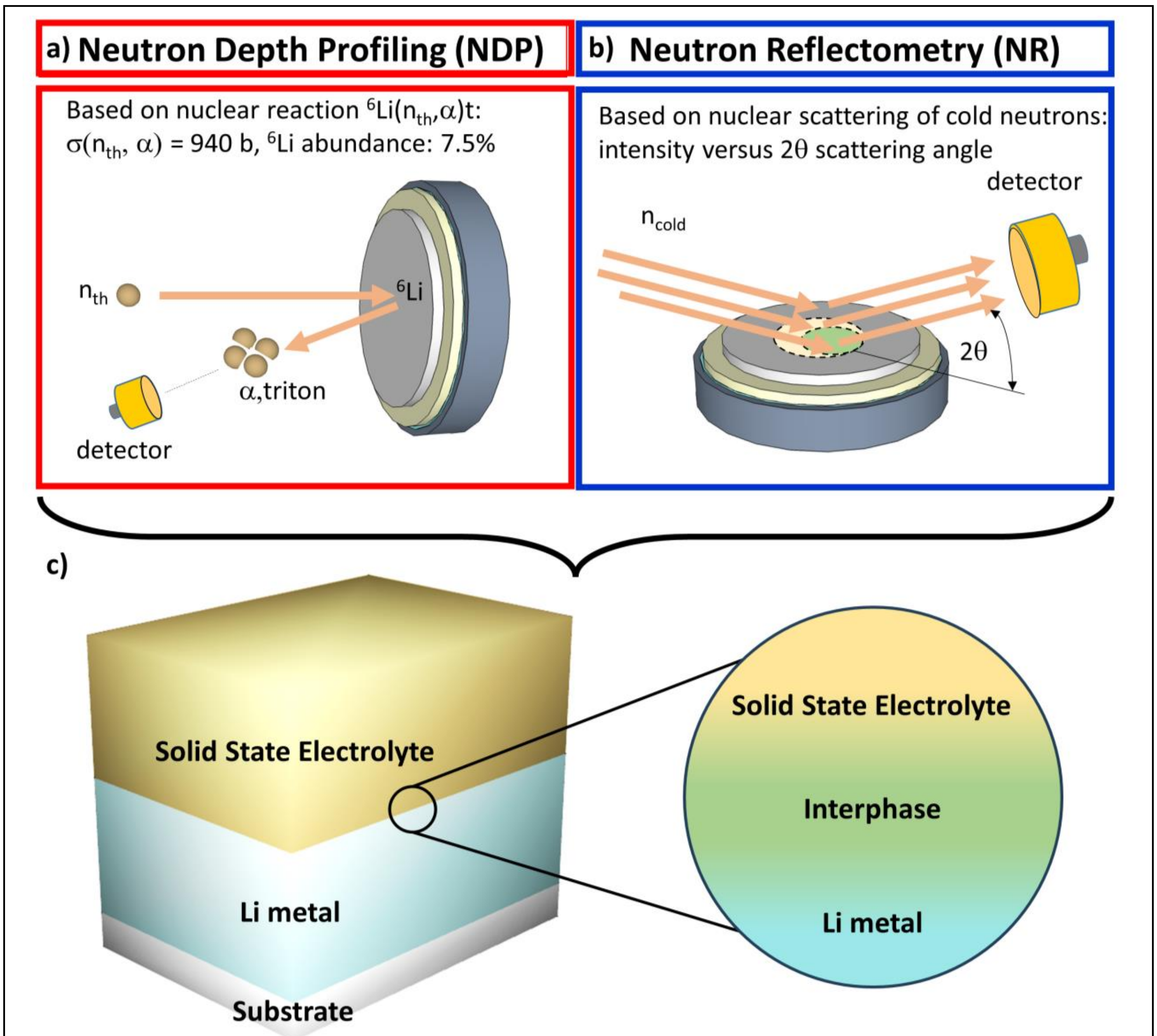

**Figure 1.** a) Schematic illustration of neutron depth profiling (NDP) where thermal neutrons ($n_{th}$) penetrate a sample and react with $Li_6$ producing both α and triton radiation which are then measured. b) Schematic illustration of neutron reflectometry (NR) where cold neutrons ($n_{cold}$) reflect and refract through a thin film multilayer forming a reflected neutron spectra with both constructive and destructive interference resulting in a Kiessig fringe pattern on the detector. c) Schematic illustration of a typical sample consisting of a solid electrolyte connected (SSE) paired with a Li metal electrode. The blowout highlights the formation of an interphase between the solid electrolyte and the Li metal which can be studied with both NDP and NR.

The samples used in both the NDP and NR measurements were made using sputtering and thermal evaporation as detailed in Table 1 and Figure S1. The sample sets for NDP contain both thin (~100 nm) and thick (~500 nm) LiPON films deposited on Li metal. Figure 2 shows the α radiation NDP data for both thin and thick LiPON films deposited onto Li metal and onto the bare substrates as controls. NDP data is taken by measuring the intensity of the radiation emitted from the sample as a function of energy. The measured intensity indicates the number α particles of a given energy hitting the detectors. The Li concentration throughout the sample is then determined by analyzing the energy and count of the α particles. The goal of this set of measurements is to 1) determine if we can resolve a distinct interphase between Li and LiPON, 2) determine the estimated thickness of the interphase, and 3) determine the resolution of NDP for distinguishing interphases between Li metal and LiPON. For all 6 NDP samples we see a distinct high intensity peak coming from the Li metal and a lower intensity shoulder coming from the LiPON. In the next several paragraphs we will analyze the intensity and energy of the Li peak and the LiPON shoulder to answer the above questions.

| Table 1 | Title | Layer 1 | Layer 2 | Layer 3 |
|---|---|---|---|---|
| NDP-1 | Thin LiPON | LiPON ~100 nm | | |
| NDP-2 | Thin LiPON-Li | Li ~500 nm | LiPON ~100 nm | |
| NDP-3 | Thin LiPON-Ni-Li | Li ~500 nm | Ni ~50 nm | LiPON ~100 nm |
| NDP-4 | Thick LiPON | LiPON ~500 nm | | |
| NDP-5 | Thick LiPON-Li | Li ~500 nm | LiPON ~500 nm | |
| NDP-6 | Thick LiPON-Ni-Li | Li ~500 nm | Ni ~50 nm | LiPON ~500 nm |
| NR-1 | LiPON-Li (NR) | Ni ~10 nm | LiPON ~200 nm | Li ~2000 nm |

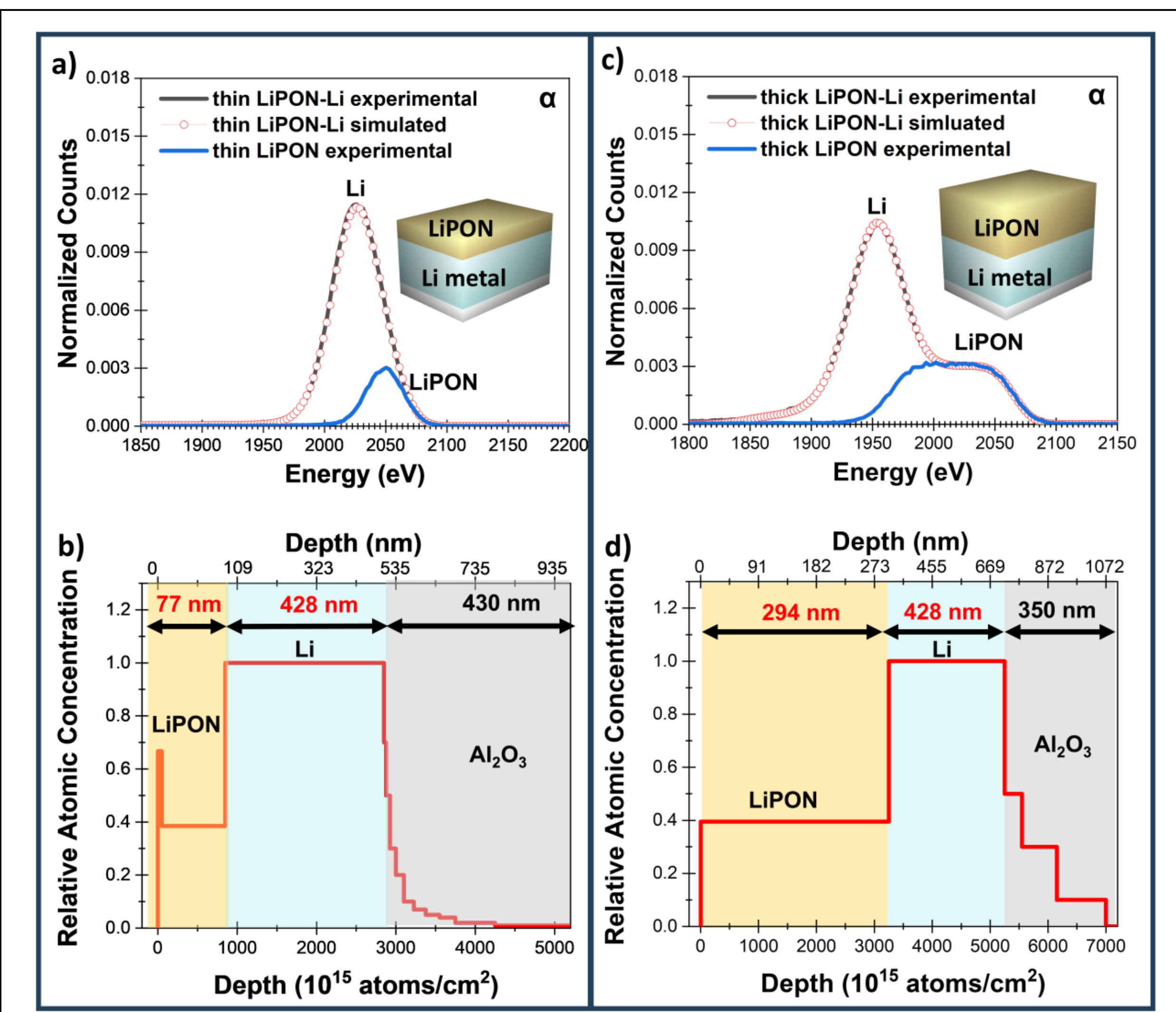

**Figure 2.** a) α radiation signal from neutron depth profiling of a sample consisting of nominally 100 nm of LiPON deposited onto 400 nm of Li metal. The black solid line shows the experimental data, and the red circles represent and simulated fit to the experimental data. The blue curve is the signal from a witness sample that only contained the 100 nm of LiPON. b) A depth representation of the simulated data in a) showing an estimated LiPON thickness of 77 nm, and a Li thickness of 428 nm. c) α radiation signal from neutron depth profiling of a sample consisting of nominally 300 nm of LiPON deposited onto 400 nm of Li metal. The black solid line shows the experimental data, and the red circles represent and simulated fit to the experimental data. The blue curve is the signal from a witness sample that only contained the 300 nm of LiPON. d) A depth representation of the simulated data in b) showing a LiPON thickness of 294 nm and a Li thickness of 428 nm.

Figure 2a shows the NDP profile from the α radiation for the thin LiPON-Li sample (black) and the accompanying thin LiPON control (blue). For this thin LiPON-Li sample there is only a single peak centered at channel 2026 eV. The profile for the control (blue) shows that the LiPON peak is centered at 2050 eV and is convoluted within the Li peak. Simulation of the data (red circles in Figure 2a and Figure 2c) estimates the LiPON layer is 77 nm thick, and the Li layer is 428 nm thick. Figure 2c shows the NDP dataset for the Li coated with the thick LiPON (black) and the associated LiPON control (blue). In this case the Li peak center is shifted from 2026 eV down to 1956 eV. A LiPON shoulder is also clearly visible from 2002-2096 eV. The peak intensity for the LiPON shoulder is ~3.4 times smaller than the Li peak as one would expect for the relative difference in Li concentration in Li metal vs. LiPON. The modeling data presented as the open circles fit in Figure 2c and in Figure 2d shows a LiPON thickness of 294 nm and a Li thickness of 428 nm. The LiPON composition was assumed to be $Li_{2.94}PO_{3.50}N_{0.31}$ determined for LiPON deposited from the same process in the same deposition chamber.[38] In both cases there is also some Li that migrates into the $Al_2O_3$ substrate. This is not surprising and not of significant importance to the questions we are trying to answer here.

| Table 2 | Sample Description | Energy Spread (eV) | Li Peak Center (eV) | Peak Intensity (a.u.) | LiPON Shoulder (eV) | LiPON shoulder Intensity (a.u.) |
|---|---|---|---|---|---|---|
| NDP-1 | Thin LiPON | 2015-2096 | --- | --- | 2015-2096 | 0.0030 |
| NDP-2 | Thin LiPON-Li | 1961-2096 | 2026 | 0.0113 | --- | --- |
| NDP-3 | Thin LiPON-Ni-Li | 1907-2096 | 1980 | 0.0087 | 2015-2096 | 0.0031 |
| NDP-4 | Thick LiPON | 1934-2096 | --- | --- | 1934-2096 | 0.0031 |
| NDP-5 | Thick LiPON-Li | 1880-2096 | 1956 | 0.0105 | 2001-2096 | 0.0031 |
| NDP-6 | Thick LiPON-Ni-Li | 1853-2096 | 1910 | 0.0083 | 1947-2096 | 0.0030 |

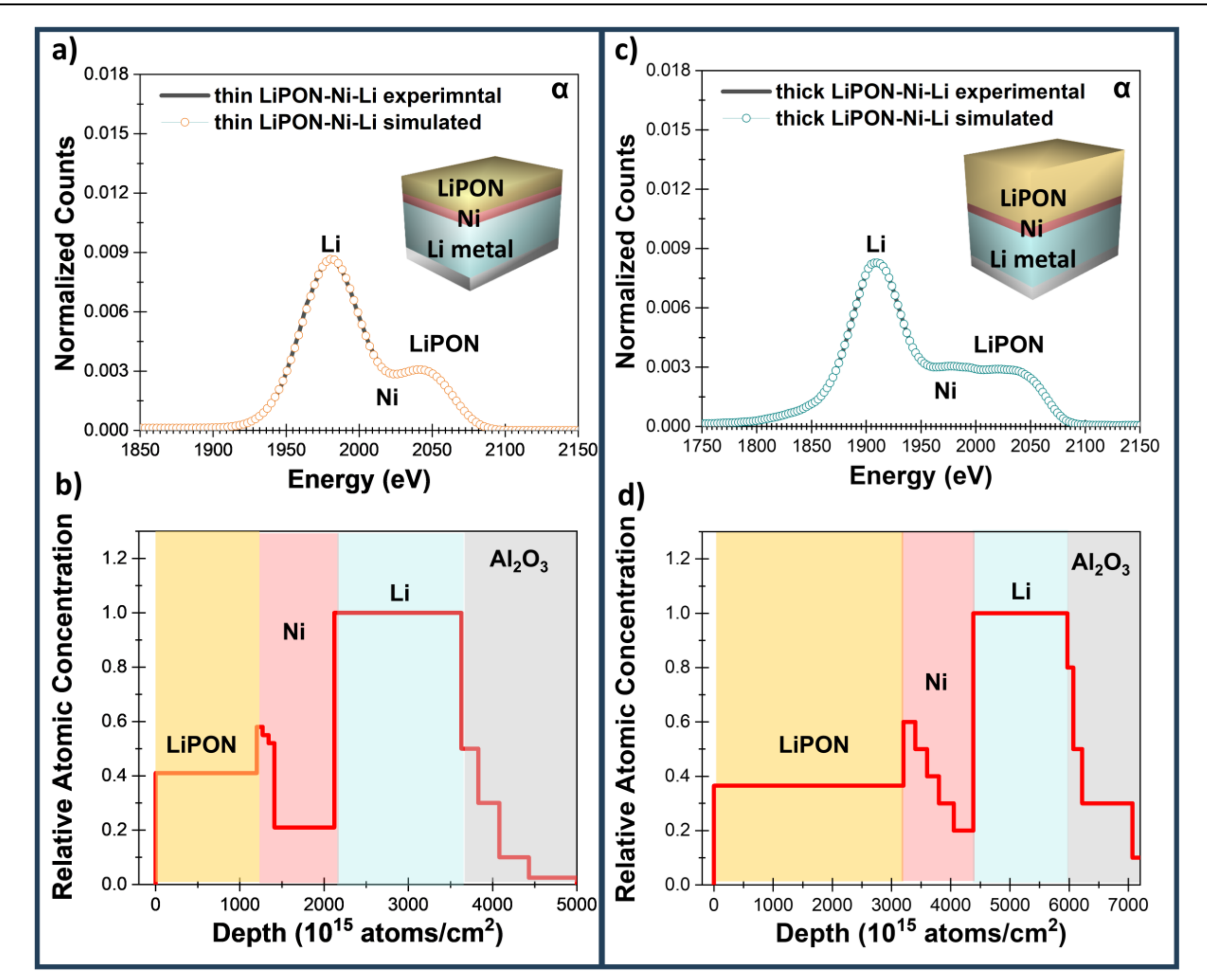

**Figure 3.** a) α radiation signal from neutron depth profiling of a sample consisting of nominally 100 nm of LiPON deposited onto ~50 nm of Ni and 400 nm of Li metal. The black solid line shows the experimental data, and the orange circles represent and simulated fit to the experimental data. b) A depth representation of the simulated data in a) represented in terms of atoms/cm$^2$ due to uncertainty in the Ni density. c) α radiation signal from neutron depth profiling of a sample consisting of nominally 300 nm of LiPON deposited onto 50 nm of Ni and 400 nm of Li metal. The black solid line shows the experimental data, and the red circles represent and simulated fit to the experimental data. d) A depth representation of the simulated data in b) represented in terms of atoms/cm$^2$ due to uncertainty in the Ni density.

Based on this NDP data we can draw several key conclusions. First, for LiPON and the Li metal to be distinguishable the solid electrolyte needs to be at least 100-200 nm thick. Second, for this data there is no distinct interphase that is observed in the NDP data as demonstrated by the simulated data which contains with no interphase, but shows an excellent fit to the LiPON data.

As the interphase that forms between LiPON and Li metal is not easily distinguishable in NDP, the natural question is what is the resolution for measuring Li/SSE interphases with NDP? Included in the answer to that question is an upper bound for the interphase that forms between LiPON and Li metal.

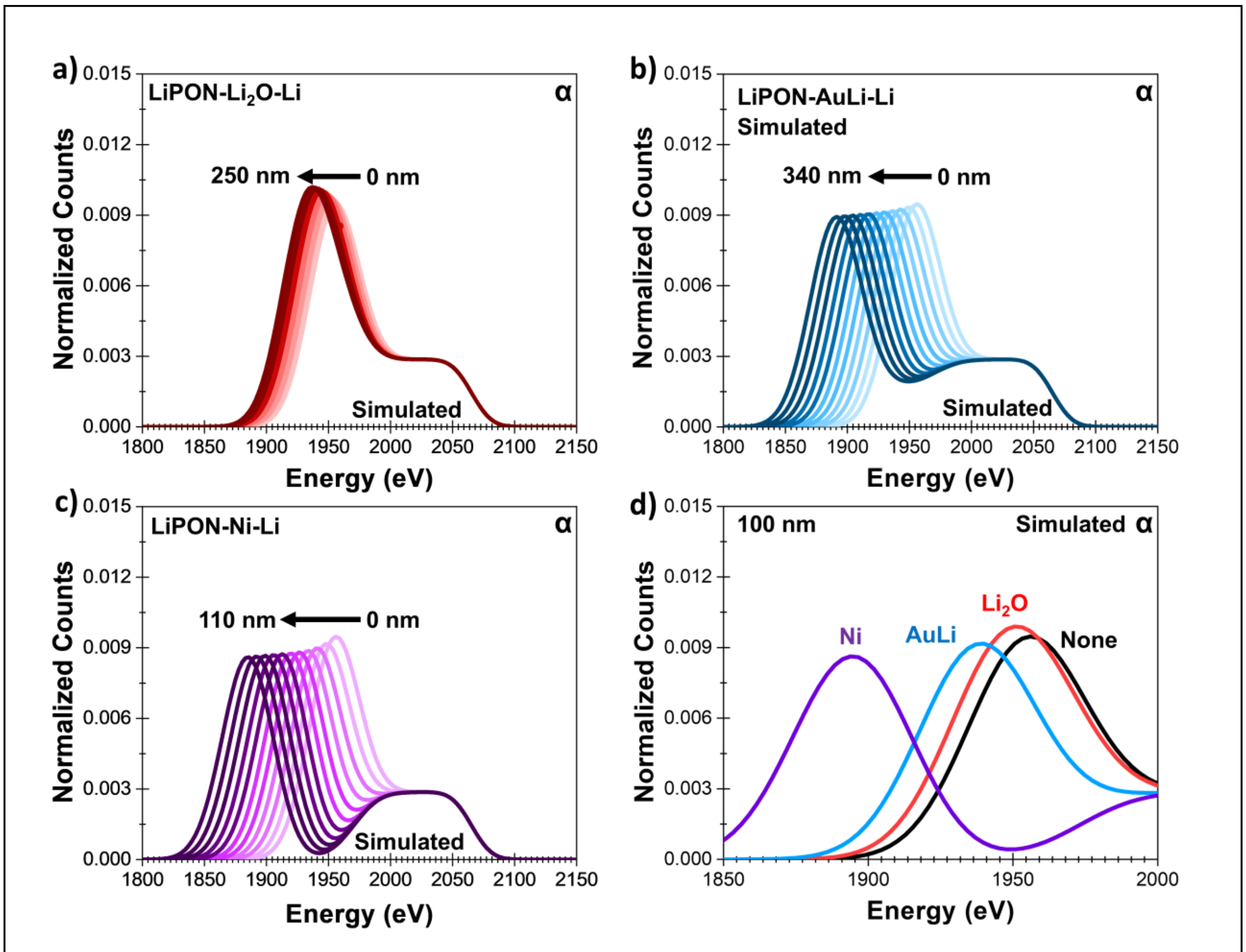

**Figure 4.** a) Simulated neutron depth profile showing the alpha radiation signal for a LiPON/Li system with a simulated $Li_2O$ interphase layer ranging in thickness from 0 to 1000 nm. b) Simulated neutron depth profile showing the alpha radiation signal for a LiPON/Li system with a simulated AuLi alloy interphase layer ranging in thickness from 0 to 1000 nm. c) Simulated neutron depth profile showing the alpha radiation signal for a LiPON/Li system with a simulated Ni metal interphase layer ranging in thickness from 0 to 1000 nm. d) simulated NDP profiles focused on the Li metal peak comparing the impact of 100 nm interlayers of $Li_2O$, AuLi, and Ni metal. All the spectra here show simulated results for α radiation.

To determine the spatial resolution of the NDP measurements we developed another set of samples with an artificial Ni based interphase between the Li metal and the LiPON. Due to the need for an air-free transfer the system used to deposit Ni was only equipped with RF magnetron sputtering with slow deposition rates. This resulted in some amount of O being included in the films enabling some Li from the LiPON deposition to react with the NiO layer forming a mixed Li-Ni-O layer. For simplicity, the labeling just uses Ni. Figure 3 shows the NDP results for both the thin LiPON-Ni-Li and thick LiPON-Ni-Li layers. For the sample with the thin LiPON layer, the Ni shifted the Li peak center to lower energy from 2026 eV (Figure 2a, Ni-free sample) to 1980 eV (Figure 3a,Ni-based sample). This had two distinct benefits, first the LiPON shoulder from 2015-2096 eV becomes distinct from the Li peak. Second there is a clear dip in the α particle intensity between the LiPON and Li peaks indicating the more Li poor NiO region. Because of the uncertainty in the density and exact composition of the NiO interlayer the thickness could not be estimated, but a model with a good fit was also produced using an estimate for the # of atoms/cm$^2$ clearly showing this distinct Ni rich interphase layer (Figure 3b). Atoms/cm$^2$ are units commonly used in interpreting NDP data. This unit is based on the $depth\ (cm) * atomic\ density\ \left(\frac{atoms}{cm^3}\right)$ to yield units of atoms/cm$^2$. This model clearly shows a LiPON layer that lasts for 0-1.2 x $10^{18}$ Li atoms/cm$^2$, a Ni based interlayer that goes from 1.2-2.12 x $10^{18}$ Li atoms/cm$^2$, and finally a Li layer that extends from 2.12-3.63 x $10^{18}$ Li atoms/cm$^2$. Based on the Quartz Crystal Microbalance (QCM) from the deposition the Ni layer is approximately 50 nm thick. For the thick LiPON-Ni-Li sample presented in Figure 3c, the Li peak center also shifts to lower energy from 1956 eV (Figure 2c) to 1910 eV (Figure 3c). The Li peak intensity also decreases from 0.0105 to 0.0083. This occurs due to Li migration into the NiO layer. Similarly, the model fitted to the experimental data shows the thicker LiPON layer from 0-3.2 x $10^{18}$ atoms/cm$^2$, followed by the Li rich NiO layer from 3.2-4.38 x $10^{18}$ atoms/cm$^2$, and with a final Li layer from 4.38-5.97 x $10^{18}$ atoms/cm$^2$.

Based on this data, the more subtle natural interphase between LiPON and Li metal is difficult to detect, but the higher atomic number (Z) of the Ni based interphase is more easily distinguished. To better understand the resolution limits for different types of interphase layers between Li metal and SSE we performed a series of simulations. In these simulated NDP data we used the thick LiPON-Li metal samples as a baseline and artificially inserted interphase layers of different composition and thicknesses ranging from 0-1000 x $10^{15}$ atoms/cm$^2$ in 100 x $10^{15}$ atoms/cm$^2$ increments. The exact thickness of each layer is material depending due to differences in density. We did this with three different interphase materials: $Li_2O$ which is the most likely biproduct of natural interphase layers with oxide electrolytes, a AuLi alloy layer representing one of the most compelling strategies for interphase stabilization,[39,40] and a Ni interphase representing a nominally completely Li free interphase layer. As each of these materials have different densities and molar masses this leads to interphase thicknesses of 0-250 nm for $Li_2O$, 0-340 nm for the AuLi alloy, and 0-110 nm for Ni metal. These cases represent some of the most interesting interlayer approaches to enable solid-state batteries. Figure 4a and Figure S3 shows the effect of the $Li_2O$ layers with different thicknesses 0 nm-250 nm. As the $Li_2O$ interlayer thickness increases the Li metal peak center shifts to lower energy from 1956 eV with no interphase to 1937 eV with the thickest 250 nm $Li_2O$ layer. For the AuLi alloy interlayer (Figure 4b), the central Li peak has a much more dramatic shift to lower energy from 1956 eV down to 1888 eV at 350 nm. A distinct dip in the intensity of the NDP profile is also observable starting with thicknesses of ~100 nm. For the Li-free Ni metal interlayer data (Figure 4c) we see a downshift in the Li peak center from 1956 eV to 1883 eV at 110 nm. The distinct dip in the data also happens more quickly with thicknesses starting at 21 nm. To show a more direct comparison Figure 4d shows the three interlayer types with thicknesses of ~100 nm compared to the Li peak with no interlayer. At 100 nm all three types of interphases are distinguishable as a shift of the Li peak to lower energy with shifts from 1956 eV with no interphase to 1951 eV, 1940 eV, and 1893 eV for $Li_2O$, LiAu, and Ni respectively. Based on these simulations NDP is unlikely to be able to resolve interphase layers <100 nm thick for the low Z compounds or ~10 nm for layers with high Z. The reason for this, is that while the NDP only detects the Li, the energy loss for the emitted radiation is larger when passing through higher Z based compounds.

While the discussion here has focused on α radiation as this is more depth sensitive, triton radiation is also produced when the neutrons penetrate through the sample. The experimental triton radiation datasets for the thin LiPON-Li and thick LiPON-Li as well as the simulated $Li_2O$ and Ni interlayer series is presented in Figure S4. In all these datasets there is a single Li peak centered at 2717 eV. The thicker LiPON shifts this slightly to 2712 eV. The $Li_2O$ series and the Ni series also do not cause a distinct layer separation but result in peak shifts to 2709 eV and 2698 eV respectively. Because the triton radiation is less sensitive to minor changes that would occur due to an interphase it is not the focus of this work. The advantage to triton radiation is that the triton radiation can cover a larger energy range accommodating the measurement of thicker samples.[12]

Before considering NR, there is one other important factor for NDP to discuss. When measuring an artificial interphase such as the Ni layer presented in Figure 3, NDP shows a distinct peak shift. But in many cases the interphase of most interest is the natural interphase between a solid electrolyte and Li metal or a cathode. This interphase occurs when the solid electrolyte is contacted with the electrode and typically occurs spontaneously or after the first few electrochemical cycles. For these natural interphases there will be a change in chemistry, but the total number of atoms will not change. Thus, the location of the Li peak will not shift down as we see with the artificial interphases in Figure 4. Rather, these naturally occurring interphases will primarily be visible as a decrease in the intensity of the Li peak coupled with a broadening of the peak as some of the Li migrates into the solid electrolytes. The lithiation of the Ni layer in Figure 3 is an excellent example of what we would expect to see when the reaction between the solid electrolyte and Li metal are more prevalent than the reaction of Li with LiPON. One way to enhance the visibility of this effect would be to use isotopes. For example, in this case, if the Li metal anode was composed entirely of $Li_6$ and the electrolyte of $Li_7$ then the reaction of $Li_6$ with the electrolyte would be significantly more visible than with naturally occurring Li like that used in this study.

Previously we performed an in-situ NR measurement of the interphase formation between LiPON and Li metal.[3] This experiment consisted of a quartz substrate with 10 nm of Ni, 200 nm of LiPON and 2 μm of Li where Li was plated onto the Ni through the LiPON. After an interphase had formed, Li was subsequently stripped, and the reflectivity measurement was performed. From this data, a gradient Li-LiPON interphase of 7 nm was identified. Building on that work here we used a similar sample stack with 200 nm of LiPON and 2 μm of Li on a Ni coated quartz substrate as a baseline to explore the sensitivity of NR to interphases of different thicknesses and to probe some of the limits of the technique in comparison to NDP. More details of the synthesis can be found in the supporting information. Figure 5 shows the reflectivity pattern from the NR measurement. Q in these measurements is the neutron scattering vector. A fit to the data is shown in black in Figure 5a. The scattering length density (SLD) plot associated with this fit is shown in Figure 5b. The data shows an initial Ni layer that was on the quartz substrate followed by a thin Li layer (6.5 nm), a LiPON layer (206 nm), and a thick semi-infinite Li layer greater than 1 μm. The Ni was deposited on the substrate for possible in-situ measurements but the sample was not used for that purpose (details in the supporting information). The thin Li layer deposited on the Ni comes from a battery-like lithiation effect that can occur during the magnetron sputtering of LiPON (see reference for more details).[41] With two distinct Li layers on either side of the LiPON we have the opportunity to use NR to compare the interphase that forms at the inner electrodeposited Li/LiPON interface and the outer vapor deposited LiPON/Li interface. For the inner electrodeposited Li/LiPON interphase, we observe a gradient interphase that extends from a depth of 10 nm to a depth of 14 nm for a total interphase of ~4 nm. This truly highlights the power NR in precisely measuring thin interphase layers. While this inner interphase is very thin, the outer interphase between the LiPON and the vapor deposited Li extends from 208 nm to 243 nm for a total thickness of 36 nm. The discrepancy between this thinner inner Li/LiPON interphase and the outer LiPON-Li interphase likely comes from a rougher outer LiPON surface that can occur during sputtering of ceramics and glasses. Indeed, atomic force microscopy (AFM) data for the control LiPON films deposited for NDP shows a roughness on the order of 10 nm (Figure S5). For the films with LiPON and Li metal the AFM data shows a roughness of 20-50 nm. Because of the convolution with the roughness, the true chemical interphase that arises from the reaction of LiPON with Li metal is more accurately reflected by the sharp inner interphase layer. This data also matches well with the prior in-situ measurement of the interphase with NR and a similar in-situ measurement using Coulombic titration.

To understand the limitations of NR in the context of a solid electrolyte and Li metal, we performed a series of simulations modifying the fit to the experimental dataset in Figure 5a with distinct $Li_2O$ interlayers added in between the LiPON and Li layers. This data is shown in Figure 5c-d. Figure 5c shows the simulated SLD patterns for $Li_2O$ thicknesses of 10, 20, 50, and 100 nm. The corresponding simulated reflectivity data are shown in Figure 6b. This data highlights significant changes in the Kiessig fringes across the Q spectrum shown. The data shows that NR effectively captures interphase layers less than 100 nm. While the focus of this work is on interphase layers, it also highlights some of the challenges that NR faces in terms of the needed sample geometry where the thin layers of interest must be less than ~400 nm thick. The roughness of the substrate or solid electrolyte layers will also convolute the interpretation of the data.

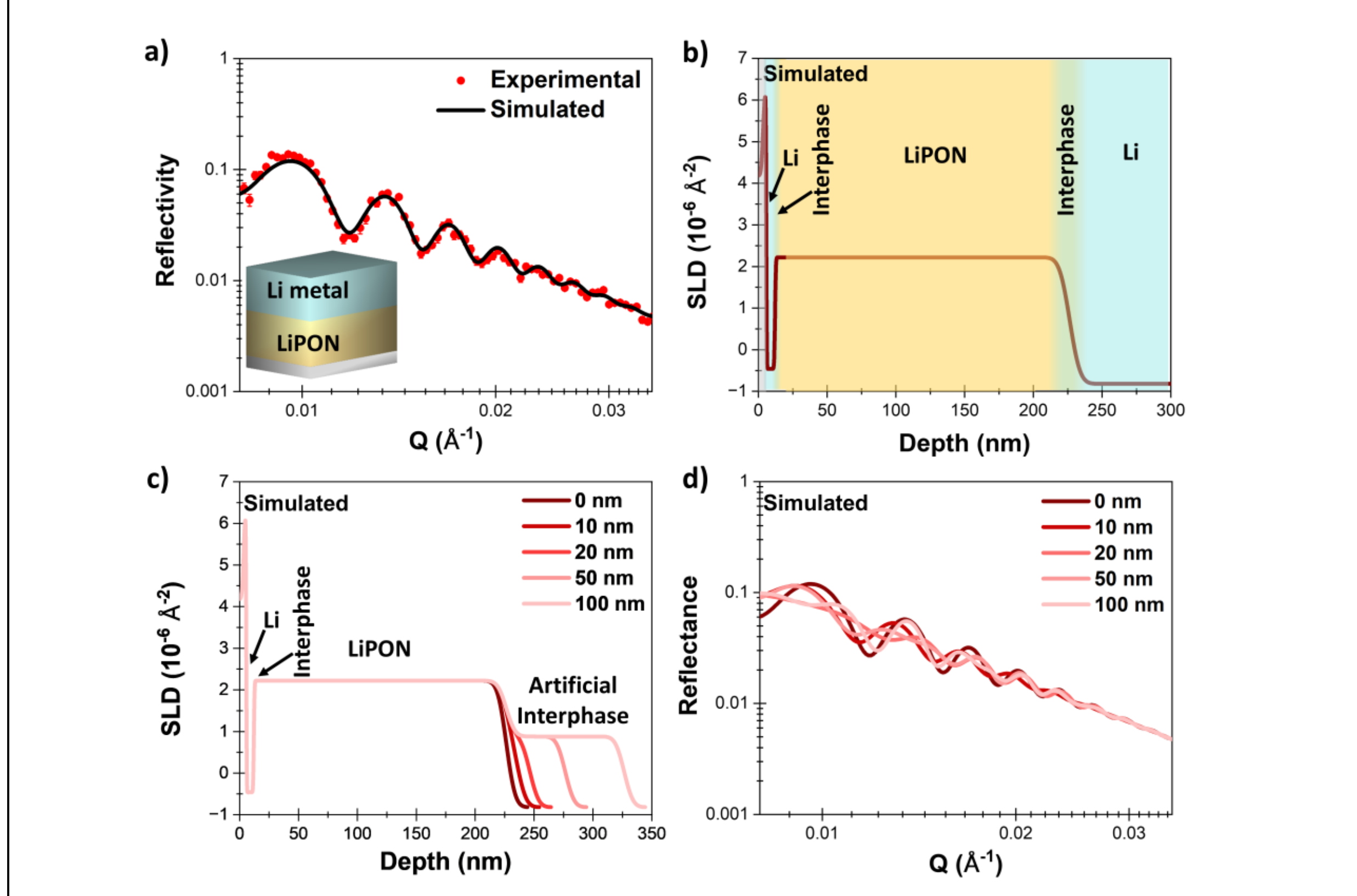

**Figure 5.** a) NR spectra shown with the reflectivity intensity for a 200 nm LiPON sample with several microns of Li metal deposited on top. The red dots are the experimental data, and the black line is the simulated fit to the experimental data. b) scattering length density (SLD) of the simulated fit to the experimental data in a) as a function of depth with the LiPON layer represented in yellow, the Li metal represented in blue, and the interphase represented in green. c) SLD as a function of depth for a simulated sample stack based on the experimental stack in a) but with an additional artificial $Li_2O$ interphase layer included between the Li and LiPON layers with thicknesses of 0, 10, 20, 50 and 100 nm. b) Simulated Kiesseg fringes for the sample series in a).

From the combined NDP and NR datasets and modeling it is clear both techniques have advantages and disadvantages that can be grouped into three categories: 1) length scale resolution (Figure 7), 2) sample requirements, and 3) chemical sensitivity. Regarding length scale resolution, NR is most effective for detecting thin interphase layers, but NDP may be better suited for thicker interphase greater than 100 nm. For NR, the roughness of the sample is the primary limitation. Practically speaking for very smooth surfaces the resolution limit is on the order of a few nanometers. For thicker interphases, NR has some significant limitations as film stacks need to be less than ~400 nm for high-quality data. In this work, we showed how a thin electrolyte film sputtered onto one of these substrates enables probing of the interphase between that solid electrolyte and Li metal. Another potential approach for ceramic or glass electrolytes includes using thick, highly polished electrolyte substrates on the order of the dimensions of the traditional NR substrates ~2" diameter and ¼" thick. Li metal could then be applied to these electrolytes by vapor deposition or melt casting. If the interphase was then less than 400 nm thick it could be resolved with NR. Another limitation of NR is that the chemical sensitivity of the technique primarily lies in the fitted values of SLD to the NR pattern. While the SLD of something like Li metal (-0.89) is relatively unique compared to almost all other compounds, the SLD of the different electrolytes and interphase layers could be similar convoluting the data and making it difficult to pinpoint exact chemical compositions. This is especially true as the SLD is also dependent on the density of the material. If the material in question is not fully dense then the SLD will be different.

For NDP, the length scale at which it excels is between 50 nm and 10 μm. The length scale resolution is also heavily dependent on whether α or triton particles are being used to determine the thickness of the interphase. For α particles, the resolution of the interphases is roughly between a few tens of nm and 5 μm. This lower limit is especially true for very thin samples. For NDP a low roughness is ideal. When using α particles the biggest limitation is that the total sample thickness needs to be less than 10 μm. If it is not, then the α and triton signals begin to overlap convoluting the sample. Triton particles on the other hand are less sensitive to very small thicknesses (Figure S4) but can be used to study films tens of microns thick when Kapton or another material is used to block the α particles. From a chemical sensitivity standpoint, NDP is specifically sensitive to Li content. This limits some broader applications such as for Na applications (here $Na_{22}$ is necessary for high detection sensitivity) but also makes the technique especially useful for Li containing samples. NDP is most powerful when significantly different quantities of Li are contained in the different layers of interest.

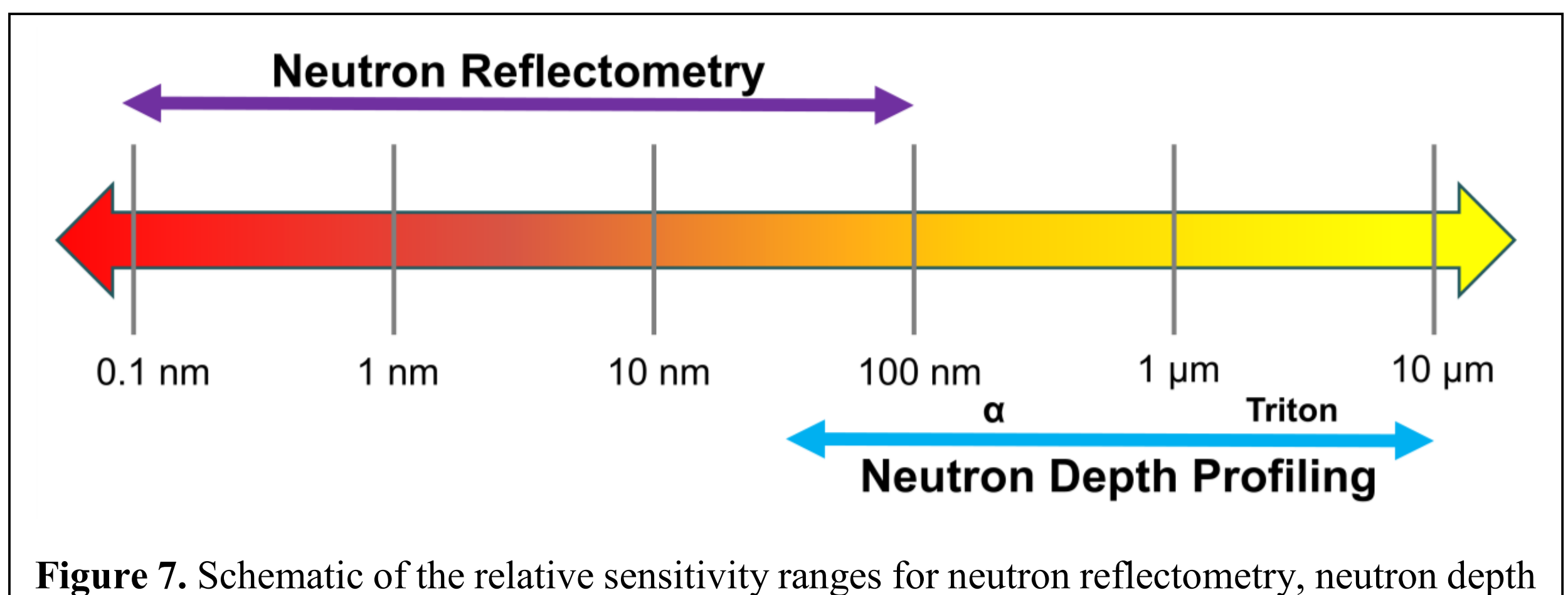

**Figure 7.** Schematic of the relative sensitivity ranges for neutron reflectometry, neutron depth profiling.

While not the focus of this paper, neutron imaging (NI) is a third technique for studying interfaces in solid-state batteries. NI is in the tomography family of techniques which enables more direct imaging of a sample. Because the minimum resolution of NI is on the order of 1 to 2 µm it is not suitable for measuring interphase between solid electrolytes and Li metal or cathode. NI can probe length scales on the order of 10 µm up to a few cm. This enables one to image an entire full cell battery stack which is not feasible with NR and NDP. Thus, NI is very well suited to observe the micron length scale morphological evolution of the interface during battery cycling.[42]

**Conclusions**

In conclusion, we have compared the use of NDP and NR to study the interphase between Li metal and the solid electrolyte LiPON both with experimental and simulated data. Both techniques confirm LiPON and Li metal form a nm thin layer, with NDP demonstrating it is less than ~100 nm thick and NR measuring a gradient interphase ~4 nm thick between electrodeposited Li and LiPON and a gradient interphase <35 nm thick between vapor deposited Li metal and LiPON. This comparative study clearly demonstrates botht he power and limitations of both NDP and NR for studying interphases in solid state batteries. NR has significantly higher resolution allowing one to measure interphase thicknesses down to the nm length scale, but is less effective at measuring thicker interphase layers. NR also comes with very stringent sample preparation requirements in terms of sample roughness and total interphase thickness. NDP on the other hand has a more limited lower limit to the resolution of ~100 nm but can measure thicker interphases. NDP samples are not quite as limited by the sample roughness and the total sample thickness can be 10s of μm. Both techniques have great promise to be powerful tools in understanding the buried solid-solid interfaces in solid state batteries.

ASSOCIATED CONTENT

**Supplementary Information**

Experimental methods, visual schematic and photographs of samples, triton radiation NDP data, atomic force microscopy of NDP samples.

AUTHOR INFORMATION

**Corresponding Authors**

Andrew S. Westover: westoveras@ornl.gov

Ralph Gilles: ralph.gilles@frm2.tum.de

ORCID:

Katie L, Browning- 0000-0001-9553-5187

Andrew S. Westover - 0000-0002-5738-1233

James F. Browning - 0000-0001-8379-259X

Mathieu Doucet – 0000-0002-5560-6478

Ralph Gilles – 0000-0003-2703-4369

**Author Contributions**

The manuscript was written through the contributions of all authors. All authors have approved the final version of the manuscript.

**Funding Sources**

The work was supported by the US-German joint collaboration on "Interfaces and Interphases In Rechargeable Li-metal based Batteries″.The work was funded by the Department of Energy’s Office of Energy Efficiency and Renewable Energy for the Vehicle Technologies Office’s US-German Cooperation on Energy Storage: Lithium-Solid-Electrolyte Interfaces program. A portion of this research used resources at the SNS, a Department of Energy (DOE) Office of Science User Facility operated by ORNL. The work presented here has received funding from the German Federal Ministry of Education and Research (BMBF) under the grant number 03XP0509C. Neutron reflectometry measurements were carried out on the Liquids Reflectometer at the SNS, which is sponsored by the Scientific User Facilities Division, Office of Basic Energy Sciences, DOE.

**Note**

This manuscript has been authored by UT-Battelle, LLC, under contract DE-AC05-00OR22725 with the US Department of Energy (DOE). The US Government retains and the publisher, by accepting the article for publication, acknowledges that the US government retains a nonexclusive, paid-up, irrevocable, worldwide license to publish or reproduce the published form of this manuscript or allow others to do so, for US government purposes. DOE will provide public access to these results of federally sponsored research in accordance with the DOE Public Access Plan (http://energy.gov/downloads/doe-public-access-plan).

ACKNOWLEDGMENT

The work was funded by the Department of Energy's Office of Energy Efficiency and Renewable Energy for the Vehicle Technologies Office's US-German Cooperation on Energy Storage: Lithium–Solid–Electrolyte Interfaces program. R. G. and N. P were funded by the Federal Ministry of Education and Research (BMBF) under grant number 03XP0509C. Funding for the NDP portion of the work was also supported by the Ministry of Education, Youth and Sports (MŠMT) of the Czech Republic within the project CZ.02.01.01/00/22_008/0004591. The authors would like to thanks Mathieu Doucet for significant contributions in performing the neutron reflectometry experiments and for performing the additional simulations of artificial interface layers for the NR data presented in Fig. 5. The authors would like to acknowledge, Tien Duong, the US program manager for this funding. The authors would also like to acknowledge Candace Halbert, who helped facilitate the Neutron Reflectometry measurements and Ethan Self and Shomaz Ul Haq for help in editing the manuscript. A portion of this research used resources at Spallation Neutron Source an OE Office of Science User Facility operated by the Oak Ridge National Laboratory. The beam time was allocated to the Liquid Reflectometer (LR) on proposal number IPTS-24417. Neutron Depth Profiling measurements were performed at the Nuclear Physics Institute (NPI) in Řež, Czech Republic within the CANAM (Center of Accelerators and Nuclear Analytical Methods) infrastructure at the LVR15 research reactor.